\documentclass[11pt,a4paper]{article}
\usepackage[utf8]{inputenc}
\usepackage{graphicx}
\usepackage{caption}
\usepackage[font=footnotesize,labelfont=bf]{caption}
\usepackage[hmargin=2.4cm,vmargin=2.8cm]{geometry}
\usepackage{amsmath}
\usepackage{authblk}
\usepackage{url}
\usepackage{xfrac}
\usepackage{multicol}
\usepackage{textgreek}
\usepackage{sectsty}
\usepackage{placeins}
\usepackage{subcaption}
\usepackage{verbatim}
\usepackage{tabularx,ragged2e,booktabs,caption}
\usepackage{amssymb}
\usepackage[overload]{empheq}
\usepackage{xcolor}
\usepackage{nccmath}
\usepackage{mathtools}
\usepackage{float}
\usepackage{titlesec}
\usepackage{enumitem}
\usepackage{wrapfig}
\usepackage{notes2bib}

\title{Average activity of excitatory and inhibitory neural populations}
\date{}
\author{Javier Roulet and Gabriel B. Mindlin}
\affil{Department of Physics, FCEyN, UBA, \\
IFIBA, CONICET\\
Ciudad Universitaria, Pab I, cp 1428, Buenos Aires, Argentina}

\begin{document}

\maketitle

%Abstract:
We develop an extension of the Ott-Antonsen method \cite{Ott2008} that allows obtaining the mean activity (spiking rate) of a population of excitable units. By means of the Ott-Antonsen method, equations for the dynamics of the order parameters of coupled excitatory and inhibitory populations of excitable units are obtained, and their mean activities are computed. Two different excitable systems are studied: Adler units and theta neurons. 
The resulting bifurcation diagrams are compared to those obtained from studying the phenomenological Wilson-Cowan model in some regions of the parameter space. Compatible behaviors, as well as higher dimensional chaotic solutions, are observed. We study numerical simulations to further validate the equations.

\bigskip

% Lead Paragraph:
\noindent\textbf{An active area of research in Physics deals with establishing connections across different scales of description for out-of-equilibrium systems. This is the reason why, for example, macroscopic models of nervous systems are usually made phenomenologically as opposed to statistically. In 2008, Ott and Antonsen developed a statistical method for obtaining the evolution of the macroscopic ``order parameter'' of a large ensemble of coupled oscillators, which describes its degree of synchronization \cite{Ott2008}. This method has recently been applied to densely connected neural populations. However, it is often the case that the mean activities of the populations (i.e. their spiking rates) are the variables of interest, particularly for behavioral control. In this paper we extend the Ott-Antonsen method, in order to obtain equations for the mean activity of a population in terms of its order parameter. We apply this result to two different models of a ``neural oscillator'', consisting of coupled excitatory and inhibitory populations of excitable units, and compare the resulting dynamics to those of the frequently used phenomenological Wilson-Cowan model. We obtain compatible behaviors in a wide range of parameter values, as well as more complex chaotic solutions.}

\section{Introduction}

The description of how thousands of fireflies, crickets or neurons fall into step, collectively synchronizing, has attracted the attention of dynamicists for decades. Winfree made significant progress in this field by arguing that in certain limits, amplitude variations could be neglected, and the oscillators could be described solely by their phases along their limit cycles \cite{Winfree1967}. Kuramoto introduced a model for a large set of weakly coupled, nearly identical oscillators, with interactions depending sinusoidally on the phase difference between each pair of units \cite{Kuramoto1975}. Interestingly, stationary solutions of this nonlinear model can be solved exactly, in the infinite-$N$ limit, with the application of self-consistency arguments \cite{Strogatz2000}. 

In 2008, Ott and Antonsen introduced an ansatz for studying the behavior of globally coupled oscillators \cite{Ott2008}, which has been most convenient for studying continuous time-dependent collective dynamics. The ansatz refers to the statistical description of the oscillators, and the result of this technique is a low dimensional system of reduced equations that describe the asymptotic behavior of the order parameter of the system. This order parameter, which is the resultant phasor of the system, describes the degree of synchrony of the ensemble.

Phase equations are not only an adequate representation for oscillatory dynamics. They can also describe the dynamics of a class of excitable systems, and large sets of coupled excitable units are a natural proxy for understanding the dynamics of many dynamically rich systems, neural networks among them. Recently, the Ott and Antonsen ansatz was used to explore the macroscopic dynamics of large ensembles of coupled excitable units \cite{Alonso2010, Alonso2011, So2014}. Yet, the macroscopic dynamics in those works was described in terms of the order parameter (as in the case of coupled oscillators), while for the study of neural arrangements, a natural macroscopic observable is the activity of the network \cite{Hoppensteadt1997, Montbrio2015}. 

In this work, we study the average activity of a large set of coupled excitable units. We are interested in a particular architecture: the neural oscillator, built out of coupled excitatory and inhibitory units. We show that the average activity of the network can be analytically computed in terms of the order parameters of the problem, and investigate the dynamics displayed by those macroscopic variables. We compare our results with the solutions of the phenomenologically derived Wilson-Cowan dynamical system. The comparison between the analytical expressions and the averages computed from numerical simulations allows us to unveil the subpopulation dynamics that coexist with different average behavior. We analyze two different cases. In the first one, the dynamics of the individual units (both excitatory and inhibitory) is modeled by Adler's equations \cite{Adler1946}. In the second case, the individual units are ``theta neurons'' \cite{Ermentrout1986}. In both cases, we emphasize the similarities and differences between the macroscopic solutions and those of the phenomenological Wilson-Cowan system.

The work is organized as follows. Section \ref{sec:Adler_units} presents the analysis of the first model, which consists of a set of impulsively coupled excitable phase oscillators, whose individual dynamics are ruled by Adler's equations. Section \ref{sec:activities} contains the analytical results for that case, which include the computation of the average activity as a function of the order parameters of the problem. Section \ref{sec:theta_neurons} presents a similar analysis for the second case under study, corresponding to the theta neurons. In Section \ref{sec:bifurcation_diagrams} we discuss the bifurcation diagrams for the order parameter equations of the two models under analysis, and we compare them with a bifurcation diagram for the Wilson Cowan model. We report regions of the parameter space where the dynamics of our macroscopic models derived from first principles is similar to those of the phenomenological Wilson-Cowan system. We also report and discuss the departures from it. Numerical simulations of extended systems are described in Section \ref{sec:numerical_simulations}. We finish with our discussion and conclusions in Section \ref{sec:conclusions}.

\section{Coupled Adler's equations} \label{sec:Adler_units}

By ``neural oscillator'' we refer to an ensemble of two large populations of globally coupled excitable units: one excitatory and the other inhibitory. The proposed dynamics for the individual units is a phase oscillator in an excitable regime. One hypothesis in this approach is that all the relevant information about the internal state of an individual unit can be contained in a phase variable $\theta$ on the unit circle. Consequently, the microscopic variables in our model are as many phases $\{\theta_i\}$ as there are units in the population. If these excitable oscillators are used to model the dynamics of neurons, then the cycles of $\theta$ are interpreted as the neuron's spikes.

One widely used model of an excitable oscillator is given by the Adler equation $\dot{\theta} = \omega_i-\cos \theta_i$ \cite{Adler1946}. It features a Saddle-Node in Limit Cycle (SNILC) bifurcation at $\omega=1$, which is the known mechanism for the onset of spiking activity in Type-I neurons. For $\omega_i\lesssim1$, the unit is said to be in the excitable regime, with a stable resting state close to an unstable one, near $\theta\approx0$. That is, perturbations to the resting state larger than a certain threshold can trigger a large reaction on the system (a ``spike''). The threshold size depends on $\omega$, which can be interpreted as the intrinsic excitability of the unit. The Adler model has another SNILC bifurcation at $\omega=-1$, where the unit starts spiking with its phase running backwards (an unwanted dynamical feature if the excitable units are asked to represent neurons).

In turn, the units are supposed to be globally coupled, so that their evolutions obey the following equations:

\begin{subequations} \label{eq:model}
	\begin{align}[left = \empheqlbrace\,]
		\dot{\theta}_i(t) &= \omega_i - \cos\theta_i(t) + I\big(\{\theta_j(t)\}, \{\tilde\theta_j(t)\}\big),\\
		\dot{\tilde{\theta}}_i(t) &= \tilde{\omega}_i - \cos\tilde{\theta}_i(t) + \tilde I\big(\{\theta_j(t)\}, \{\tilde\theta_j(t)\}\big),
	\end{align}
\end{subequations}

\noindent where the untilded variables refer to units in the excitatory population, and tilded variables (\textasciitilde) to the inhibitory ones. The first two terms in the Eqs.~\eqref{eq:model} describe the internal dynamics of each unit in the neural oscillator, and the couplings $I, \tilde I$ parametrize the interaction between them. These are chosen to be:

\begin{subequations} \label{eq:couplings}
	\begin{align}
		I\big(\{\theta_j\}, \{\tilde\theta_j\}\big) &= \frac{k_E}{N} \sum_{j=1}^{N} (1-\cos\theta_j) - \frac{k_I}{\tilde{N}} \sum_{j=1}^{\tilde{N}} (1-\cos\tilde{\theta}_j),\\
		\tilde I\big(\{\theta_j\}, \{\tilde\theta_j\}\big) &= \frac{\tilde{k}_E}{N} \sum_{j=1}^{N} (1-\cos\theta_j) - \frac{\tilde{k}_I}{\tilde{N}} \sum_{j=1}^{\tilde{N}} (1-\cos\tilde{\theta}_j),
	\end{align}
\end{subequations}

 \noindent where the different $k$, $\tilde k>0$ describe the coupling strengths between neurons, and $N$, $\tilde N$ are the number of neurons in each of the two populations. The functional form is chosen so that the $j$-th unit influences the others via an impulsive term proportional to $(1-\cos\theta_j)$. This term is maximum at $\theta_j=\pi$ (where the spike occurs), and nearly zero close to the resting state $\theta_j\approx0$. It differs from the Kuramoto coupling $\sin(\theta_i-\theta_j)$ in that it only depends on the phase of the pre-synaptic unit, and it is always excitatory for excitatory units (and inhibitory for inhibitory units). This is represented by the sign accompanying each term, which determines whether the phases of the post-synaptic units are driven towards or away from threshold from the resting states.

A first macroscopic variable describing the collective behavior of the system, the Kuramoto order parameter, can be defined for each of the two sub-populations, averaging their phasors:

\begin{subequations} \label{eq:z_discrete}
	\begin{align}
		z(t) &= \frac{1}{N}\sum_{j=1}^{N} e^{i\theta_j(t)}, \\
		\tilde{z}(t) &= \frac{1}{\tilde{N}}\sum_{j=1}^{\tilde{N}} e^{i\tilde{\theta}_j(t)}. 
	\end{align}
\end{subequations}

These variables account for the synchrony within the sub-populations: if all the oscillations are in phase, these order parameters will present moduli equal to one. On the other hand, if the populations are active, i.e. with units spiking, but out of synchrony, the order parameters will present small values. In this sense, the order parameters do not capture all the features of a macroscopic state.

These order parameters allow us to rewrite Eq.~\eqref{eq:couplings} in a compact way:

\begin{subequations} \label{eq:I}
	\begin{align}
		I\big(z,\tilde z\big) &= k_E \left(1-\text{Re\,} z\right)-k_I \left(1-\text{Re\,}\tilde{z}\right),\\
		\tilde I\big(z,\tilde z\big) &= \tilde k_E \left(1-\text{Re\,} z\right)-\tilde{k}_I \left(1-\text{Re\,}\tilde{z}\right),
	\end{align}
\end{subequations}

\noindent which makes the system in Eq.~\eqref{eq:model} suitable to the application of the method introduced in 2008 by Ott and Antonsen \cite{Ott2008}. The essence of the computation consists in describing the state of each population of neurons through a distribution function, expanding it in Fourier modes, and using a continuity equation to derive the dynamical rules governing their behavior. Following their procedure, we first approximate the problem assuming an infinite population. The system description is now made in terms of the distributions $f(\theta,\omega,t)$, $\tilde f(\theta,\omega,t)$ that represent the density of units with a given excitability $\omega$ and phase $\theta$ in the excitatory and inhibitory populations respectively. In the following discussion, we show the steps of our calculation for the excitatory population, and a completely analogous procedure is carried out for the inhibitory one.

The distribution functions are normalized so that

\begin{equation*} \label{eq:norm_f}
	\int_{-\infty}^\infty {\rm d}\omega \int_0^{2\pi} {\rm d}\theta\, f(\theta,\omega,t) = 1,
\end{equation*}

\noindent with the excitabilities distributed according to

\begin{equation} \label{eq:g}
	 g(\omega)=\int_0^{2\pi} f(\theta,\omega,t) {\rm d}\theta,
\end{equation}

\noindent which is time-independent since the excitabilities are assumed to be constant. In this representation, the order parameters will be expressed as integrals, namely:

\begin{equation*}
		z(t) =\int_{-\infty}^\infty {\rm d}\omega \int_0^{2\pi} {\rm d}\theta\, f(\theta,\omega,t)e^{i\theta},
\end{equation*}

\noindent and our macroscopic questions can be addressed as we compute 
the distributions $f$, $\tilde{f}$.

Conservation of neurons with excitability $\omega$ means that these satisfy the continuity equation

\begin{equation} \label{eq:continuity}
	\frac{\partial f}{\partial t} + \frac{\partial}{\partial \theta}(f v) = 0,
\end{equation}

\noindent where the velocity is:

\begin{equation} \label{eq:v}
	v(\theta,\omega,t) = \omega - \cos\theta + I\big(z(t),\tilde z(t)\big).
\end{equation}

One way to solve this problem consists of performing a mode decomposition of the distributions, and finding the dynamics of the mode amplitudes. By virtue of Eq.~\eqref{eq:g}, $f$ can be decomposed as:

\begin{equation} \label{eq:f}
	f(\theta,\omega,t) = \frac{g(\omega)}{2\pi} \bigg[ 1 + \Big( \sum_{n\geqslant1} \alpha_n(\omega,t) e^{-in\theta} + \text{c.c.} \Big) \bigg],
\end{equation}

\noindent where c.c. means the complex conjugate of the preceding term. In principle, substitution of Eq.~\eqref{eq:f} in Eq.~\eqref{eq:continuity} leads to an infinite set of equations for the evolution of each $\alpha_n$. Yet, Ott and Antonsen found an ansatz (OA ansatz) that simplifies the problem \cite{Ott2008}: by proposing

\begin{equation} \label{eq:ansatzOA}
	\alpha_n(\omega,t) = \big[\alpha(\omega,t)\big]^n,
\end{equation}

\noindent the equations for all the modes are satisfied as long as the first mode satisfies:

\begin{equation} \label{eq:dalphadt}
	\frac{\partial \alpha}{\partial t} = i \big(\omega + I(z,\tilde z)\big)\alpha - \mfrac{i}{2} \left(1+\alpha^2\right).
\end{equation}

The relation between $\alpha$ and $z$ is obtained by multiplying Eq.~\eqref{eq:f} by $e^{i\theta}$ and integrating in $\theta$ and $\omega$:

\begin{equation} \label{eq:z_int_galpha}
	\int_{-\infty}^\infty g(\omega)\alpha(\omega,t) {\rm d}\omega = z(t)
\end{equation}

\noindent ($\alpha(\omega,t)$ can be interpreted as an order parameter restricted to the units with excitability $\omega$). To solve Eq.~\eqref{eq:dalphadt} we still have to compute the integral in Eq.~\eqref{eq:z_int_galpha}, which requires assuming a specific distribution $g(\omega)$ for the system's excitabilities. The Lorentzian distribution is particularly useful here. It is defined as

\begin{equation} \label{eq:lorentzian}
	g(\omega)=\frac{\Delta}{\pi}\frac{1}{(\omega-\omega_0)^2+\Delta^2},
\end{equation}

\noindent which has a maximum at $\omega_0$ and a half-width at half-maximum $\Delta$. Setting $g(\omega)$ as in Eq.~\eqref{eq:lorentzian} and assuming that $\alpha(\omega,t)$ is analytic in the complex $\omega$ upper half-plane, we can solve Eq.~\eqref{eq:z_int_galpha} by contour integration, evaluating $\alpha$ at the pole $\omega_0+i\Delta$:

\begin{equation} \label{eq:z_alpha}
	\alpha(\omega_0+i\Delta,t)=z(t).
\end{equation}

Evaluating Eq.~\eqref{eq:dalphadt} at the pole, the partial differential equations for the first mode amplitudes $\alpha$ and $\tilde{\alpha}$ become a coupled, 4-dimensional dynamical system of first order ordinary differential equations for the order parameters, namely:

\begin{subequations} \label{eq:dzdt}
	\begin{align}[left = \empheqlbrace\,]
		\dot{z} &= \big[-\Delta + i\big(\omega_0 + I(z,\tilde z)\big)\big]z - \mfrac{i}{2}\big(1+z^2\big), \\
		\dot{\tilde z} &= \big[-\tilde\Delta + i\big(\tilde\omega_0 + \tilde I (z, \tilde z)\big)\big]\tilde z - \mfrac{i}{2}\big(1+\tilde z^2\big).
	\end{align}
\end{subequations}

Ott and Antonsen demonstrated that the ansatz Eq.~\eqref{eq:ansatzOA} defines an invariant manifold, which is globally attracting for the order parameters under very general conditions \cite{Ott2009}. 
In this way, Eq.~\eqref{eq:dzdt} describes the long term solution of the problem, regardless of the initial conditions.

\section{Computation of the average activity} \label{sec:activities}

As we discussed in the previous section, the order parameters $z$ and $\tilde{z}$ describe the degree of synchrony of the system. Another sensible description of its macroscopic behavior is the level of activity of the sub-populations, understood as the total number of spikes taking place, per unit of time. This quantity can be computed as the flux of phasors through $\theta=\pi$:

\begin{subequations} \label{eq:phi}
	\begin{align}
		\phi(t) &= \int_{-\infty}^\infty f(\theta,\omega,t) v(\theta,\omega,t)\Big|_{\theta=\pi} {\rm d}\omega,\\
		\tilde{\phi}(t) &= \int_{-\infty}^\infty \tilde{f}(\theta,\omega,t) \tilde{v}(\theta,\omega,t)\Big|_{\theta=\pi} {\rm d}\omega,
	\end{align}
\end{subequations}

\noindent with $v,\tilde{v}$ satisfying Eq.~\eqref{eq:v}. A convenient expression for $f(\pi,\omega,t)$ can be obtained by imposing the OA ansatz Eq.~\eqref{eq:ansatzOA} explicitly in Eq.~\eqref{eq:f}. Now each sum becomes a geometric series that can be written in terms of $\alpha$:

\begin{equation*}
	\sum_{n\geqslant1} \alpha^n e^{-i\pi n}=\sum_{n\geqslant1} (-\alpha)^n = \frac{1}{1+\alpha}-1.
\end{equation*}

The expression for $f$ given in Eq.~\eqref{eq:f} is not analytical when extended to the complex plane, because of the appearance of $\alpha^*$ in the complex conjugate term, and therefore Eq.~\eqref{eq:phi} cannot be integrated by means of the residue theorem. In order to solve this problem we propose decomposing $f$ into two terms, one of them analytical, and the other its complex conjugate. This yields:

\begin{equation} \label{eq:f_OA}
	f(\pi,\omega,t) = \frac{g(\omega)}{2\pi} \left( \frac{1}{1+\alpha(\omega,t)}-\frac{1}{2} \right) + \text{c.c.}
\end{equation}

\begin{equation} \label{eq:act}
	\phi(t) =\int_{-\infty}^\infty \frac{g(\omega)}{2\pi} \left( \frac{1}{1+\alpha(\omega,t)}-\frac{1}{2} \right) v(\pi,\omega,t) {\rm d}\omega + \text{c.c.},
\end{equation}

\noindent since both $g$ and $v$ are real for real $\omega$, and therefore equal to their complex conjugates. The integrand in Eq.~\eqref{eq:act} is now analytical and we can apply the residue theorem.

This integral needs to be evaluated in principal value, as $v(\omega)\sim\omega$ for large $\omega$, causing the integral to diverge in $\pm\infty$. The infinite contribution to the mean activity made by the ``unphysical" units with $\omega\to\infty$ is canceled by the negative infinite activity of the equally unphysical units with $\omega\to-\infty$, leaving only the contribution of the (``physical") intermediate $\omega$ units. In Appendix \ref{app:divergences} we give a rigorous method for avoiding these infinities by slightly changing the distribution function $g(\omega)$.

We can perform the integral in Eq.~\eqref{eq:act} by means of the residue theorem. To do so, we enclose the upper complex half-plane with a semicircle of radius $R\to\infty$ and subtract its contribution to the integral, which yields:

\begin{equation*}
		\phi(t) = \bigg[ \frac{1}{2\pi} \left( \frac{1}{1+z(t)}-\frac{1}{2} \right) v(\omega_0+i\Delta,\pi,t) + \lim_{R\to\infty} \frac{i\Delta}{2\pi^2}\int_0^\pi \left( \frac{1}{\alpha(Re^{i\varphi},t)+1}-\frac{1}{2} \right){\rm d}\varphi \bigg]+\text{c.c.},
\end{equation*}

\noindent where we made use of Eq.~\eqref{eq:z_alpha} to write the first term as a function of the order parameter. Using that $\alpha(\omega,t>0)\to0$ as $\text{Im\,}\omega\to\infty$ (which follows from Eq.~\eqref{eq:dalphadt}) we can perform the integral in the second term. This leads to the important result:

\begin{subequations} \label{eq:activity}
	\begin{align}
		\phi(z,\tilde z)=&\frac{1}{\pi}\bigg(\frac{1+\text{Re\,} z}{|1+z|^2}-\frac{1}{2}\bigg) \Big(\omega_0+1+I(z,\tilde z)\Big) + \frac{\Delta}{\pi}\frac{\text{Im\,} z}{|1+z|^2}, \\
		\tilde{\phi}(z, \tilde z) =& \frac{1}{\pi} \bigg(\frac{1+\text{Re\,} \tilde z}{|1+\tilde z|^2}-\frac{1}{2}\bigg) \Big(\tilde{\omega}_0+1+\tilde I(z, \tilde z)\Big) + \frac{\tilde\Delta}{\pi}\frac{\text{Im\,}\tilde z}{|1+\tilde z|^2}. 
	\end{align}
\end{subequations}

Eqs.~\eqref{eq:activity} give an explicit relation for the mean activities in terms of the order parameters. They were obtained under the OA ansatz without making any additional assumption. In this way, once we compute the order parameters satisfying our nonlinear ordinary differential equations \eqref{eq:dzdt}, we can obtain the activity of each sub-population by the evaluation of the algebraic expression above.

This result allows to make a connection between the order-parameter-based description of the neural oscillator suggested by the Ott-Antonsen statistical method, and the mean-activity-based description made by phenomenological, additive models (as the Wilson and Cowan neural oscillator model). A similar search for a macroscopic description of coupled excitable cells in terms of activity was carried out recently by Montbri\'o et al. in \cite{Montbrio2015}. Another derivation of activity rates in a population of theta neurons, valid for order parameters constant in time, was done in \cite{Laing2014}. 

Note that the mean activities obtained from Eq.~\eqref{eq:activity} are a projection of the 4-dimensional dynamics given by Eq.~\eqref{eq:dzdt}, while the Wilson-Cowan model obeys a simpler, 2-dimensional system. Thus, their qualitative behaviors can only be compatible if a further dimensional collapse occurs.

\section{Results for the theta neuron model} \label{sec:theta_neurons}

A word of caution should be said about the consequences that the Lorentzian distribution proposal has on the mean activity defined by Eq.~\eqref{eq:phi}. In general, the single-unit models subjected to the Ott-Antonsen method are meaningful in a range of parameter values (physical regime), but feature some kind of pathological behavior when the frequency parameter becomes large, like the arbitrarily fast or ``backward'' spikes in the Adler model presented above. However, the Ott-Antonsen prescription requires integrating a nonzero (Lorentzian) distribution over the whole infinite range of frequencies. It is difficult to prevent the single-unit models from having some unphysical regime at big parameter values (for example, by changing the phasor velocities' dependence with $\omega$), because our prescription for computing the mean activities requires that $g\cdot v$ be analytical and integrable in the whole upper complex $\omega$ plane. Thus, we can try to overcome this problem by choosing a narrow distribution width $\Delta$, so that the vast majority of the units lie in the physical regime. Indeed, the impact of the unphysical units in the order parameter dynamics is limited, since the influence $\sim(k/N)(1-\cos\theta)$ that each neuron has on the others is bounded by $2k/N$, independently of the spiking frequency. In this way, ``a small proportion of unphysical units" means that they make a small contribution to the system dynamics. However, we defined the mean activity as the spiking frequency of the population, to which the unphysical units can significantly contribute. We have shown in Appendix \ref{app:divergences} the mechanism by which the problem resolves in the Adler-units model, which involves the cancellation of opposite diverging contributions made by the high- and low-$\omega$ tails of the Lorentzian distribution.

It is worth exploring a model for which no parameter value places the units into a backwards oscillation. This can be achieved if the individual units are ``theta neurons", a canonical model for Type-I excitability neurons \cite{Ermentrout1986}. The proposed dynamics for the individual units is also a phase oscillator in an excitable regime, but the equation driving its dynamics is given by:

\begin{equation}
 \dot{\theta}_i(t) = 1-\cos\theta_i(t) + (1+\cos\theta_i(t)) \eta_i
\end{equation}

\noindent where now $\eta$ plays the role of the frequency parameter. As in the previous model, the system undergoes a saddle node in a limit cycle, a global bifurcation leading the unit from an excitable regime to an oscillatory one, at $\eta=0$. The main difference between this model and the previous one is that now no parameter value puts our units into a backwards oscillation. This automatically resolves the divergence in the negative $\eta$ end of the integral in Eq.~\eqref{eq:phi}. The spiking frequency of the individual (uncoupled) neurons still diverges for large $\eta$, but now only as $\sqrt\eta$ instead of linearly: $\tau(\eta)=\int_0^{2\pi} \dot\theta^{-1} {\rm d}\theta=\pi/\sqrt\eta$ for $\eta>0$. The $\eta^{-2}$ decay of the Lorentzian function is then sufficient to render the integral finite.

The equations now read:

\begin{subequations} \label{eq:model_tn} % tn = theta neuron
	\begin{align}[left = \empheqlbrace\,]
		\dot{\theta}_i(t) &= 1-\cos\theta_i(t) + (1+\cos\theta_i(t)) \Big[\eta_i + I\big(\{\theta_j(t)\}, \{\tilde\theta_j(t)\}\big)\Big],\\
		\dot{\tilde\theta}_i(t) &= 1-\cos\tilde\theta_i(t) + (1+\cos\tilde\theta_i(t)) \Big[\tilde\eta_i + \tilde I\big(\{\theta_j(t)\}, \{\tilde\theta_j(t)\}\big)\Big],
	\end{align}
\end{subequations}

\noindent with $I$, $\tilde I$ given by Eq.~\eqref{eq:I}.

A procedure analogous to the one followed in Section \ref{sec:Adler_units} leads to the following equations for the order parameters:

\begin{subequations} \label{eq:dzdt_tn}
	\begin{align}[left = \empheqlbrace\,]
		\dot z &= 2iz + \frac{1}{2}\Big[-\Delta + i\big(\eta_0 - 1 + I(z, \tilde z)\big)\Big](z+1)^2, \label{eq:dzdtE_tn}\\
		\dot {\tilde z} &= 2i\tilde z + \frac{1}{2}\Big[-\tilde\Delta + i\big(\tilde\eta_0 - 1 + \tilde I(z,\tilde z) \big)\Big](\tilde z+1)^2. \label{eq:dzdtI_tn}
	\end{align}
\end{subequations}

The computation of the activity for the populations can be carried out following the steps described in Section \ref{sec:activities}, and we obtain:

\begin{subequations} \label{eq:activity_tn}
	\begin{align}
		\phi(z) = \frac{2}{\pi}\left(\frac{1+\text{Re\,} z}{|1+z|^2}-\frac{1}{2} \right)\\
		\tilde\phi(\tilde z) = \frac{2}{\pi}\left(\frac{1+\text{Re\,} \tilde z}{|1+\tilde z|^2}-\frac{1}{2} \right)
	\end{align}
\end{subequations}

Notably, in this case each sub-population's activity is determined solely by its order parameter (i.e. it does not depend on the order parameter of the other sub-population).

\section{Bifurcation diagrams} \label{sec:bifurcation_diagrams}

In this section we compare a bifurcation diagram of the Wilson-Cowan neural oscillator, to diagrams obtained for the coupled Adler units and for the coupled theta neurons. The local bifurcations were computed numerically with \textit{PyDSTool} \cite{Clewley2012}.

The Wilson-Cowan oscillator is a phenomenologically derived model for the activity of two coupled neural populations, excitatory and inhibitory. The variables $x$ and $y$ represent their activities, and their dynamics are prescribed by the following differential equations:

\begin{subequations} \label{eq:WCmodel}
	\begin{align}[left = \empheqlbrace\,]
		\dot{x} = -x+S(\rho_x + ax-by),\\
		\dot{y} = -y+S(\rho_y + cx-dy),
	\end{align}
\end{subequations}

\noindent where $S(\xi)=1/(1+e^{-\xi})$ is a sigmoidal function that represents the nonlinear nature of the response: beyond some input level, the average activity of a population no longer increases its value. The excitatory or inhibitory nature of $x$ and $y$ are represented by the sign accompanying the positive coupling parameters $a$, $b$, $c$, $d$. The parameters $\rho_x$, $\rho_y$ describe the external inputs to the excitatory and inhibitory populations respectively. For instance, these could be from other regions of the nervous system, and are expected to be the most dynamical system parameters. Therefore, it is useful to understand the bifurcation diagram for the parameters $\rho_x$ and $\rho_y$.

\begin{figure}
	\centering
	\includegraphics[width=\linewidth]{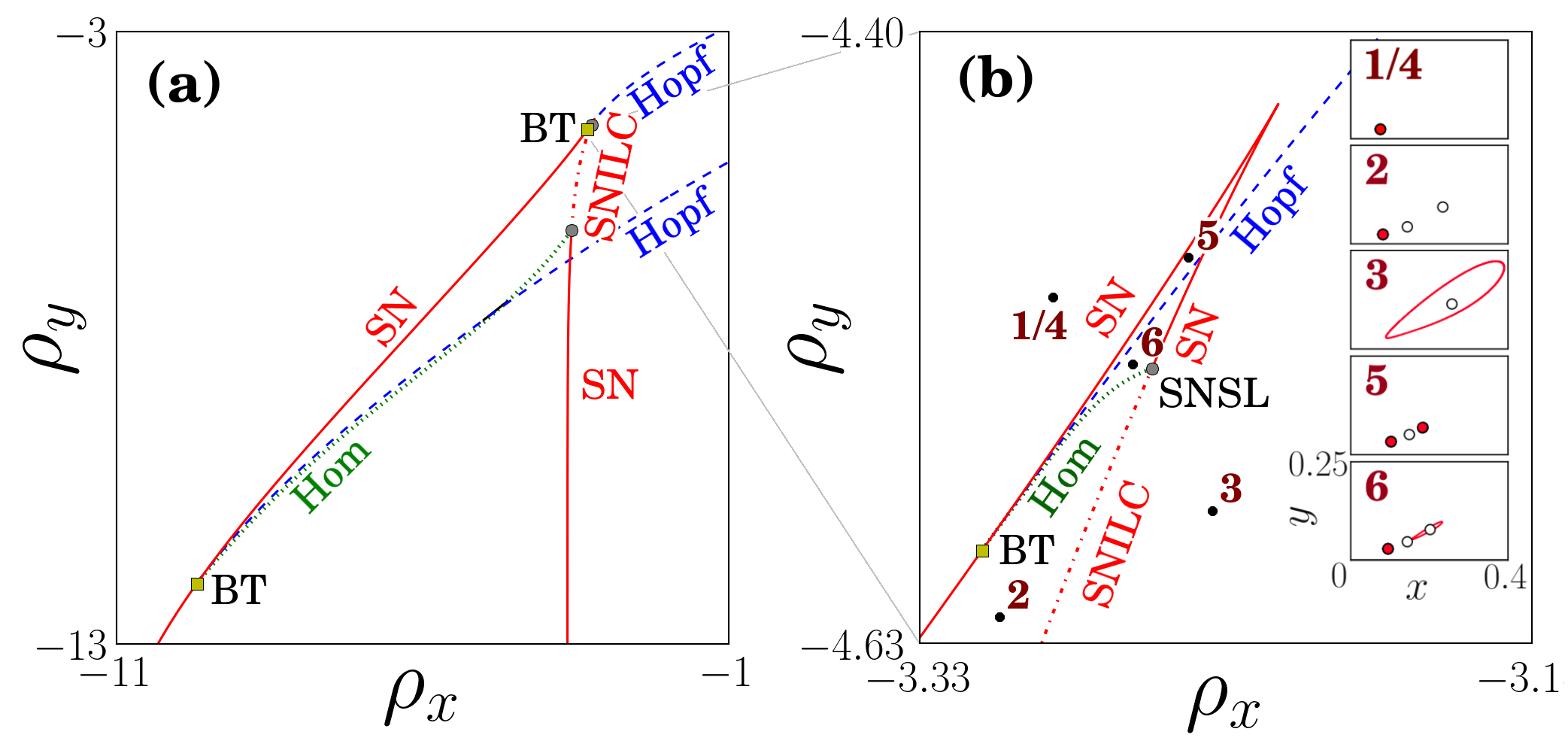}
	\caption{Bifurcation diagram of the Wilson-Cowan model (Eq.~\eqref{eq:WCmodel}) for the parameters $\rho_x$ and $\rho_y$. The remaining parameters have been set to $a=15$, $b=15$, $c=12$, $d=5$. The right panel (b) shows a detail near one of the Bogdanov-Takens bifurcations. The bifurcation curves define 5 regions with qualitatively different limit sets. The insets show limit trajectories in the phase space $(x, y)$ at parameter values representative of each region, labeled 1-6: filled dots represent stable fixed points, empty dots unstable fixed points and closed curves represent limit cycles. Regions 1 and 4 have been identified to help later comparison to the other models.}
	\label{fig:bif_WC}
\end{figure}

The region of the parameter space that we chose to display in Fig.~\ref{fig:bif_WC} presents a variety of dynamical regimes. At the blue dashed curves labeled as ``Hopf", oscillations are born in Hopf bifurcations. The curves labeled ``SN" correspond to saddle node bifurcations, where saddles and attractors (or repulsors) meet and disappear. ``SNILC" denotes the saddle-node in a limit cycle bifurcation. In this case, before the disappearance of a saddle and a node, the unstable manifold of the saddle was part of the stable manifold of the attractor. In this way, at the bifurcation the manifolds become a limit cycle, and an infinite period bifurcation of finite (nonzero) amplitude is born. The juncture of two SN curves is known as ``cusp", and these two colliding SN curves delimit a region in the parameter space where three fixed points exist (labeled 5, 6 and 2 in Fig.~\ref{fig:bif_WC}b). The global description of the stationary dynamics is the following: increasing $\rho_x$, the attractor with small $x$ value and a saddle collide, and the only surviving attractor is one with high $x$ value: the system has been excited (regions 3 and 4, the right side of $1/4$). Analogously, decreasing $\rho_x$ leads to the disappearance of the excited attractor and the surviving attractor is the ``off, non excited" state (region 1, the left part of $1/4$). In between Hopf bifurcation lines (regions 3 and 6), the system displays stable oscillations: the excitatory and inhibitory populations get to be sequentially excited. The dotted green line, labeled ``Hom'', corresponds to a homoclinic bifurcation, at which a limit cycle collides with the saddle of the system. This curve is born tangent to a SN curve at a point where a Hopf and a SN curves touch, a codimension-2 Bogdanov-Takens bifurcation (``BT''), and dies non tangent at another SN curve, in a Saddle-Node in Saddle-Loop (SNSL) codimension-2 bifurcation. 

\begin{figure}
	\centering
	\begin{minipage}{.49\textwidth}
		\centering
		\includegraphics[width=\linewidth]{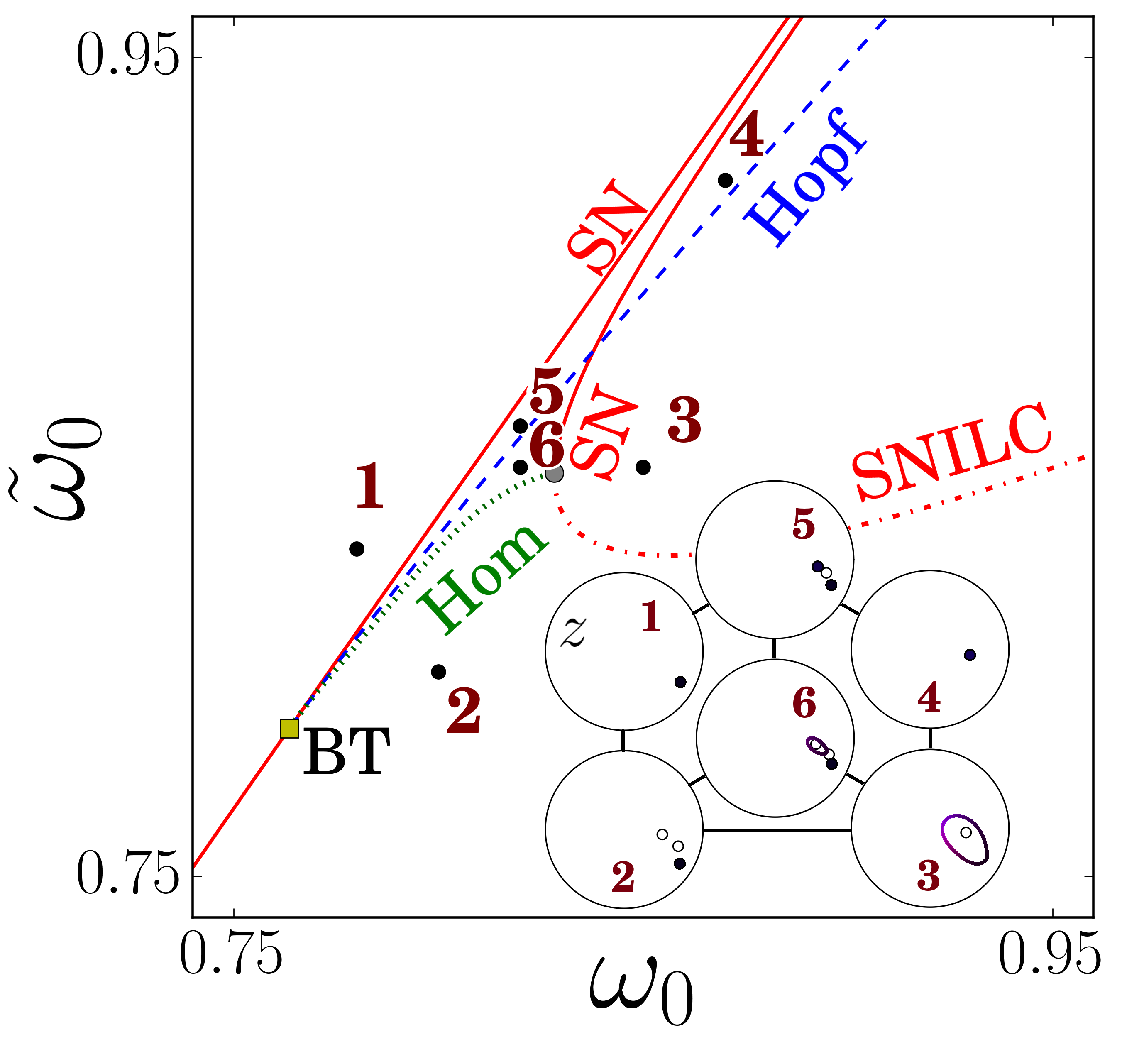}
		\captionof{figure}{Bifurcation diagram of the Adler-units model Eq.~\eqref{eq:dzdt} for the parameters $\omega_0, \tilde\omega_0$. The remaining 6 parameters were set to $\Delta = 0.1,$ $\tilde\Delta = 0.11$, $k_E = 3.0$, $\tilde k_E = 2.7$, $k_I = 2.45$, $\tilde k_I = 2.35$. The insets show the $z$ component of the limit trajectories in the $z,\tilde z$ space for a representative set $(\omega_0, \tilde\omega_0)$ in each region.}
		\label{fig:bif_Adler}
	\end{minipage}
	\quad
	\begin{minipage}{.475\textwidth}
		\centering
		\includegraphics[width=\linewidth]{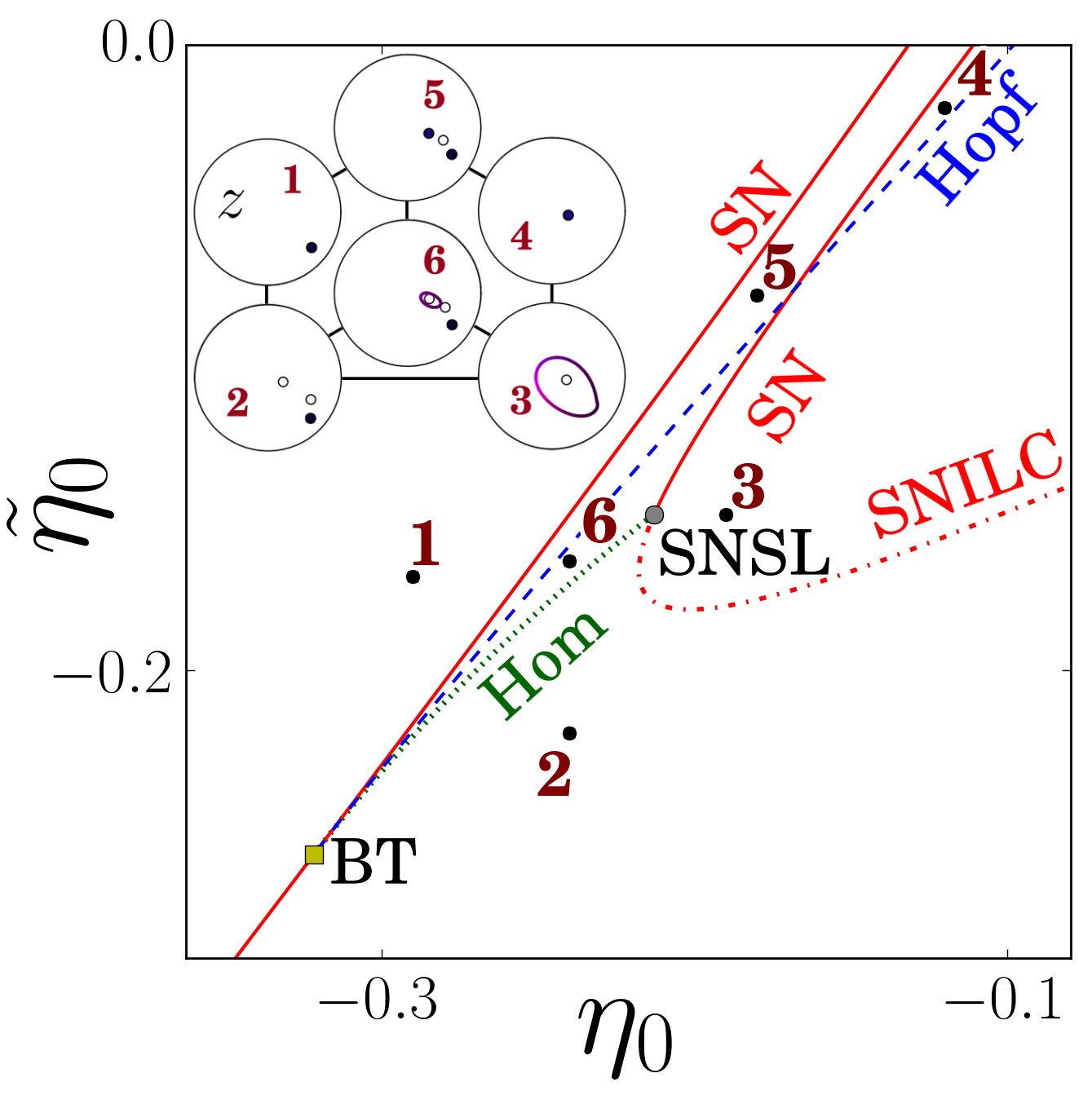}
		\captionof{figure}{Bifurcation diagram of the theta neuron model (Eq.~\eqref{eq:dzdt_tn}) for the parameters $\eta_0$, $\tilde\eta_0$. The remaining 6 parameters were chosen as in Fig.~\ref{fig:bif_Adler}.}
		\label{fig:bif_tn}
	\vspace{1.3cm}
	\end{minipage}
\end{figure}

 Fig.~\ref{fig:bif_Adler} displays the bifurcation diagram that we obtained studying the Eqs.~\eqref{eq:dzdt} for the order parameters of the Adler system, at the parameter values that we report in the caption. Each inset represents the $|z|\leqslant1$ disk, where the limit sets of the system in Eq.~\eqref{eq:dzdt} at different parameter values are projected. Fig.~\ref{fig:bif_tn} displays a similar diagram, for the equations of the theta neuron model. Both diagrams show a direct correspondence in their bifurcations and the limit sets in each region. The two SN curves approach significantly, which suggests the presence of a cusp bifurcation (an $8-2=6$ dimensional hypersurface in the 8-dimensional parameter space) nearby. However, these curves do not intersect, rather, they approach and then separate (not shown), which means that the cusp does not cross the hyperplane defined by our parameter choice. We conjecture that the regions 1 and 4 are actually the same region at the two sides of the postulated cusp (i.e., in the full parameter space, they could be connected without crossing any bifurcation). Then, these bifurcation diagrams also match the Wilson-Cowan one, displayed in Fig.~\ref{fig:bif_WC}b. They both share with the Wilson-Cowan model the coexistence of ``on" and ``off" stationary attracting states separated by a saddle, the existence of simple oscillations where the activity of the competing populations alternate, and a series of global bifurcations that allow the possibility of oscillations with critical slowing down.

Recently, complex motor patterns in birdsong production were described as the solutions displayed by a neural oscillator at the expiratory related area of the song system, when driven by inputs from other neural structures \cite{Alonso2015}. In that model, the key dynamical feature that allows reproducing those patterns was associated with the proximity between a SN and a Hopf curve, a dynamical scenario present in the Wilson-Cowan, and shared by the solutions of the average equations derived from first principles.

\begin{figure}
	\centering
	\includegraphics[width=.4\linewidth]{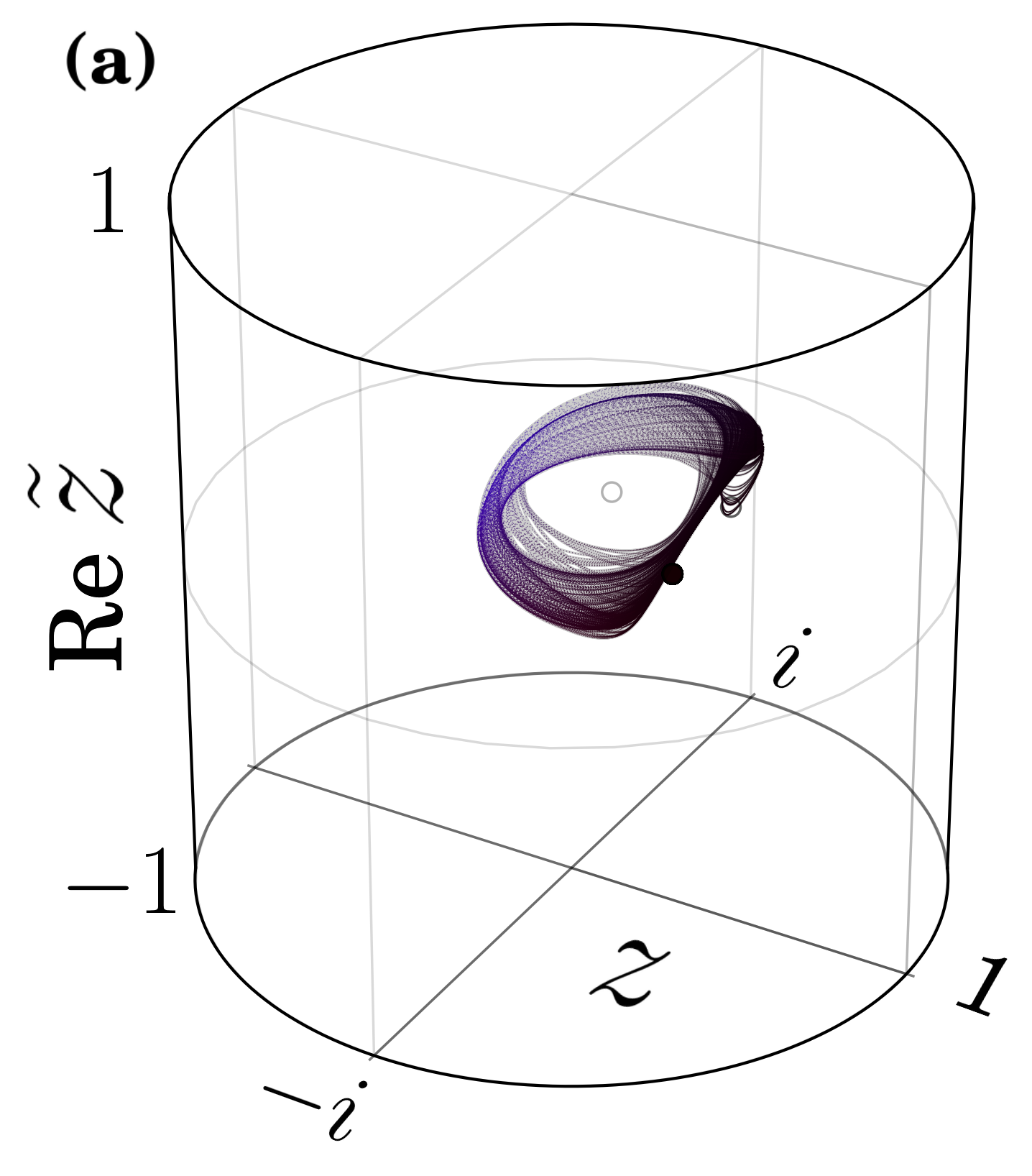}
	\qquad
	\includegraphics[width=.4\linewidth]{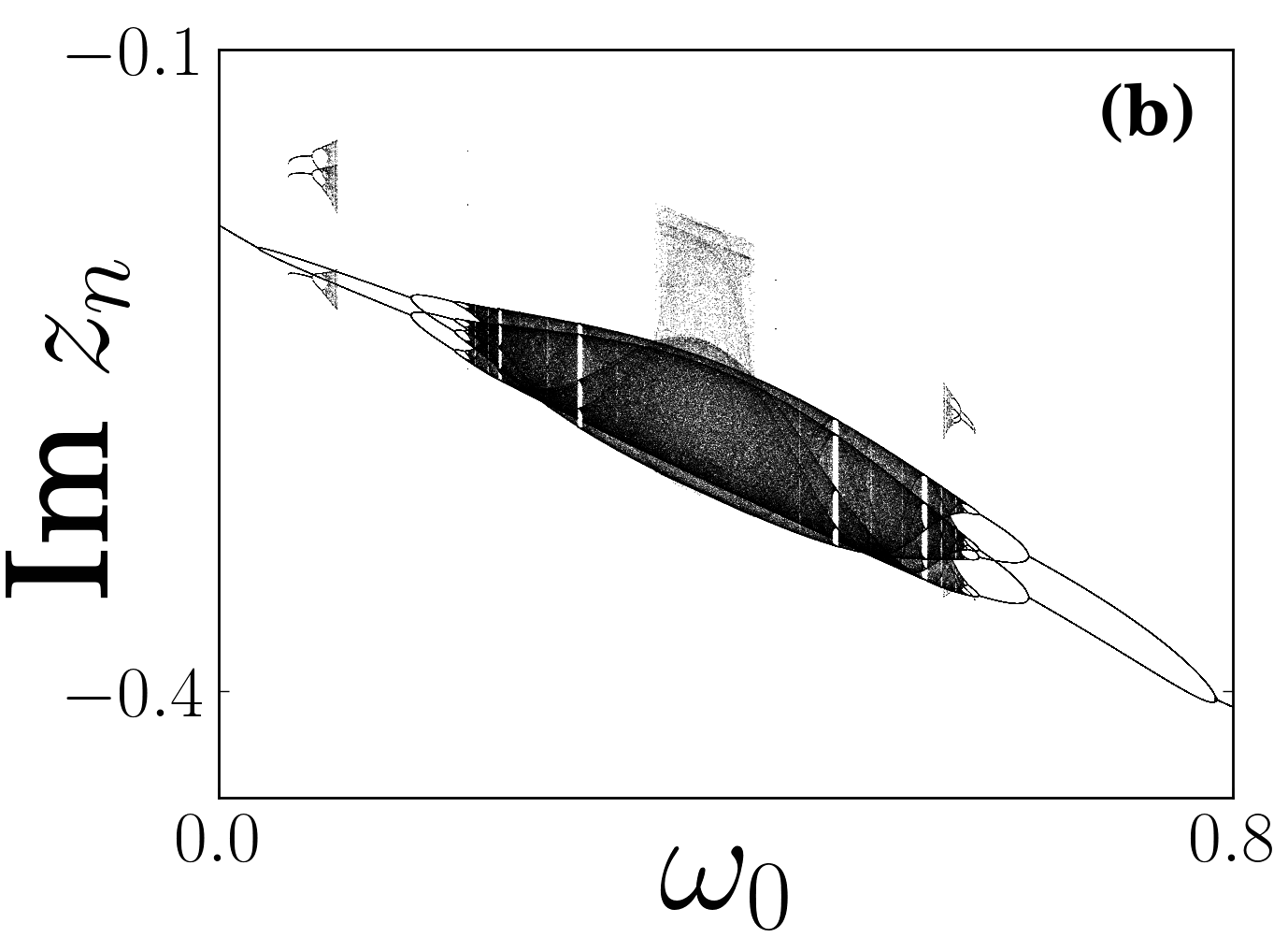}
	\caption{\textbf{(a)} Projection of a chaotic attractor for the order parameters of the Adler units system, at parameters $\omega_0 = 0.35$, $\tilde\omega_0 = 1.73$, $\Delta = 0.15$, $\tilde\Delta = 0.27$, $k_E = 9.0$, $\tilde k_E = 5.0$, $k_I = 3.5$, $\tilde k_I = 2.5$. Stable and unstable fixed points coexist, plotted as filled or empty circles respectively. \textbf{(b)} Bifurcation diagram showing the birth of the chaotic attractor as the parameters are changed. In the vertical axis we plot the imaginary part of the solution for $z$ at the intersections $z_n$ with a Poincar\'e section at angle $\psi=-\pi/2$. At $\omega_0=0$ (and all the other parameters constant) the system has a period 1 limit cycle, seen as a single intersection. Increasing $\omega_0$ continuously brings it through a cascade of period-doubling bifurcations that gives birth to the strange attractor, and a crisis in which it becomes more complex. Further increase of $\omega_0$ brings the attractor back to a simple limit cycle through the same changes in reverse order.}
	\label{fig:chaotic_Adler}
\end{figure}

Since the equations for the order parameters (in the two models analyzed in this work) are four dimensional, it is possible to find behaviors more complicated than the one present in the Wilson-Cowan model. As an example, Fig.~\ref{fig:chaotic_Adler} displays a chaotic solution. However, in wide regions of the bio-physically relevant parameter space volume, the system's attractors are those characteristic of a two dimensional dynamical system: fixed points or simple, ``untwisted'' limit cycles. In the Adler-units model, this behavior was observed in about 95\% of 6500 runs, varying independently all 8 parameters in Eqs.~\eqref{eq:dzdt} in the ranges: $\Delta\in[0.05, 0.27]$, $\omega_0\in[0.6, 1.5]$, $k\in[2, 9.5]$ for each population.

This suggests the existence of an attracting, invariant two dimensional manifold within the four dimensional phase space for those parameter values. Indeed, it is possible to find it analytically in the special case where the two populations have symmetric parameters (i.e. $\omega_0=\tilde\omega_0$, $\Delta=\tilde\Delta$, $k_E=\tilde k_E$, $k_I=\tilde k_I$). In Appendix \ref{app:2D} we show that for this specific case, the plane manifold $z=\tilde z$ is invariant and stable. We expect that departing away from the symmetric case starts causing a deformation of the two dimensional manifold before the system explores the full dimensionality. Notice that the parameters of the bifurcation diagrams in Figs.~\ref{fig:bif_Adler} and \ref{fig:bif_tn} are close to fulfilling the symmetry condition, and the system displays rich two dimensional behavior. The parameters of Fig.~\ref{fig:chaotic_Adler}, on the other hand, are not, and the system can explore a higher dimensionality.

\section{Numerical simulations} \label{sec:numerical_simulations}

In this section we analyze simulations of the full system Eq.~\eqref{eq:model}, for a network of $10^4$ Adler units in each of the two populations. These simulations allow to test the validity of the mean field Eqs.~\eqref{eq:dzdt} and \eqref{eq:activity}, for the dynamics of the order parameters and the mean activity respectively. Moreover, they help us gaining further insight into the role that the synchronization of units at the microscopic level plays on the macroscopic dynamics.

The order parameters of the simulated populations can be computed from the individual phases by means of Eq.~\eqref{eq:z_discrete}. The accuracy of the mean field Eq.~\eqref{eq:dzdt} can be tested by comparing the simulated and predicted trajectories in the $(z, \tilde z)$ space. Furthermore, the mean activity can be computed in the simulation by definition, as the fraction of phasors that crossed $\theta=\pi$ in a small time interval $\delta t$, divided by $\delta t$. Eq.~\eqref{eq:activity} makes a testable prediction of these mean activities from the order parameters of the simulation.

In the simulations, the sets of individual excitabilities and initial phases $\{\omega_i, \theta_i\}$, $\{\tilde\omega_i, \tilde\theta_i\}$ were chosen so that the resulting distribution functions satisfied the Ott-Antonsen ansatz (i.e., were given by Eq.~\eqref{eq:f_OA}) and had an arbitrary initial condition for their order parameters. This was achieved by proposing an uncorrelated initial distribution function $f(\theta,\omega,0)=g(\omega)h(\theta)$, with $g(\omega)$ given by Eq.~\eqref{eq:lorentzian} and 

\begin{equation*} \label{eq:h}
	h(\theta)=\frac{1}{2\pi}\Big(\frac{1}{1-z(0) e^{-i\theta}}-\frac{1}{2}\Big)+\text{c.c.},
\end{equation*}

\noindent which fulfills both conditions automatically. The initial phases were chosen randomly according to $h(\theta)$, and the excitabilities were generated by taking $\omega_i=\omega_0+\Delta\tan x_i$ with $\{x_i\}$ distributed uniformly in the interval $(-\pi/2,\pi/2)$. This yields the desired Lorentzian distribution for the $\{\omega_i\}$. A similar procedure was applied to the inhibitory population.

The differential equations \eqref{eq:model} and \eqref{eq:dzdt} were integrated numerically using the order 4 Runge-Kutta method with a time step of size $0.01$. A time interval of $\delta t=0.8$ was used to compute the mean activities from definition as described above.

\begin{figure}
	\centering
	\includegraphics[width=.3\linewidth]{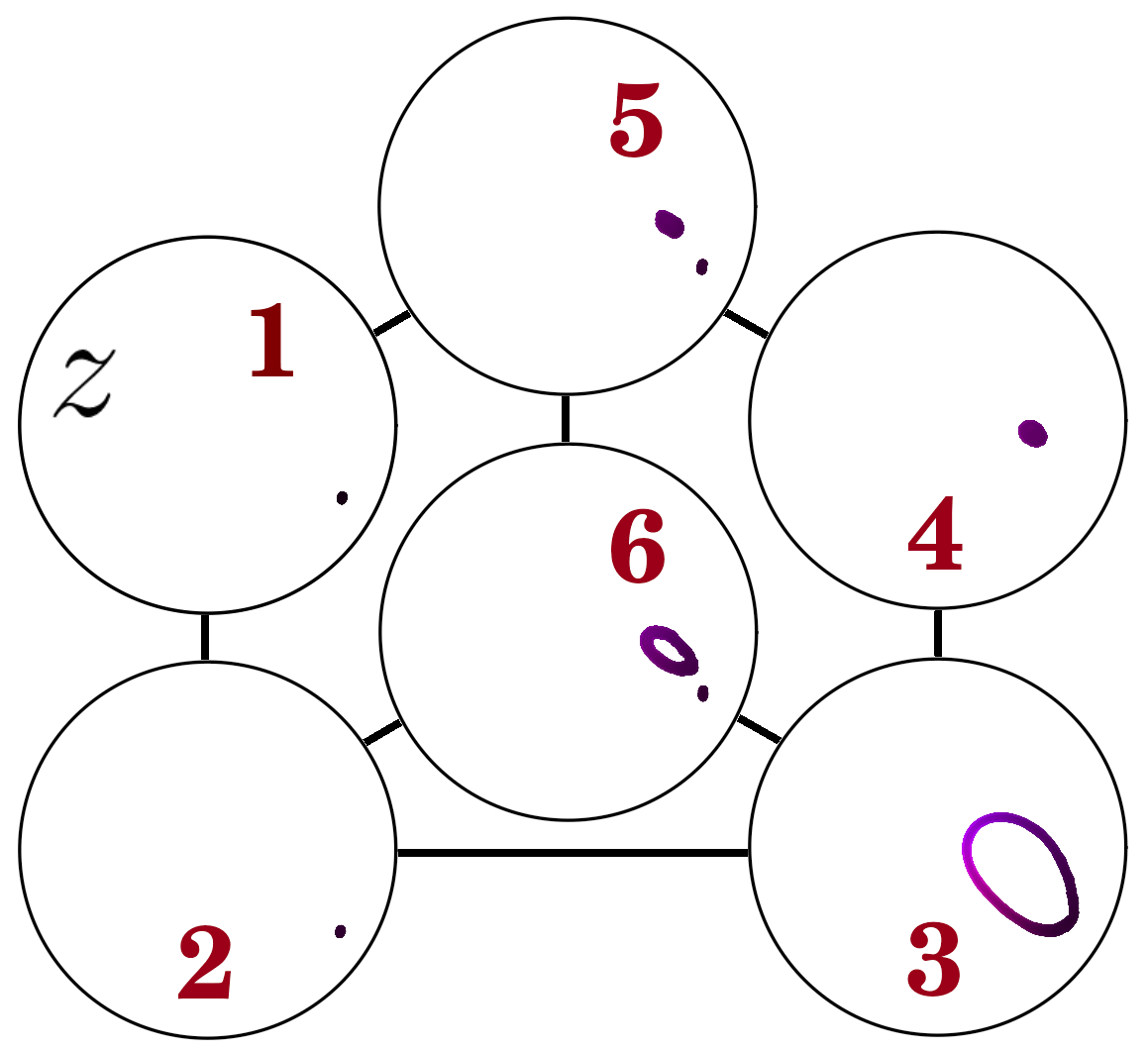}
	\caption{$z$ component of the attractor sets for the order parameters in a simulation of coupled Adler units ($10^4$ in each population), at the parameter values labeled 1-6 in Fig.~\ref{fig:bif_Adler}.}
	\label{fig:sim_attractors}
\end{figure}

\begin{figure}
	\centering
	\includegraphics[width=.65\linewidth]{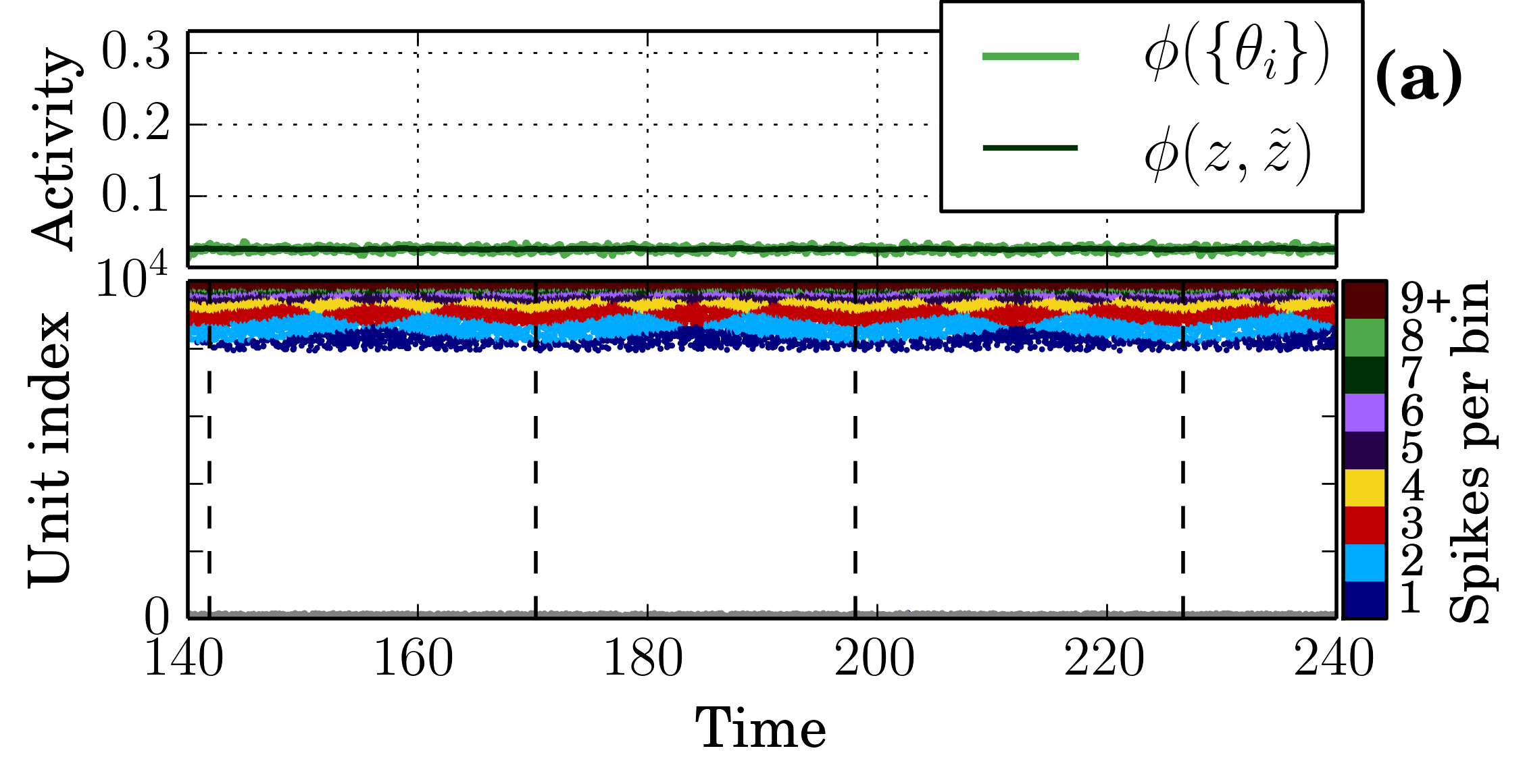}
	
	\includegraphics[width=.65\linewidth]{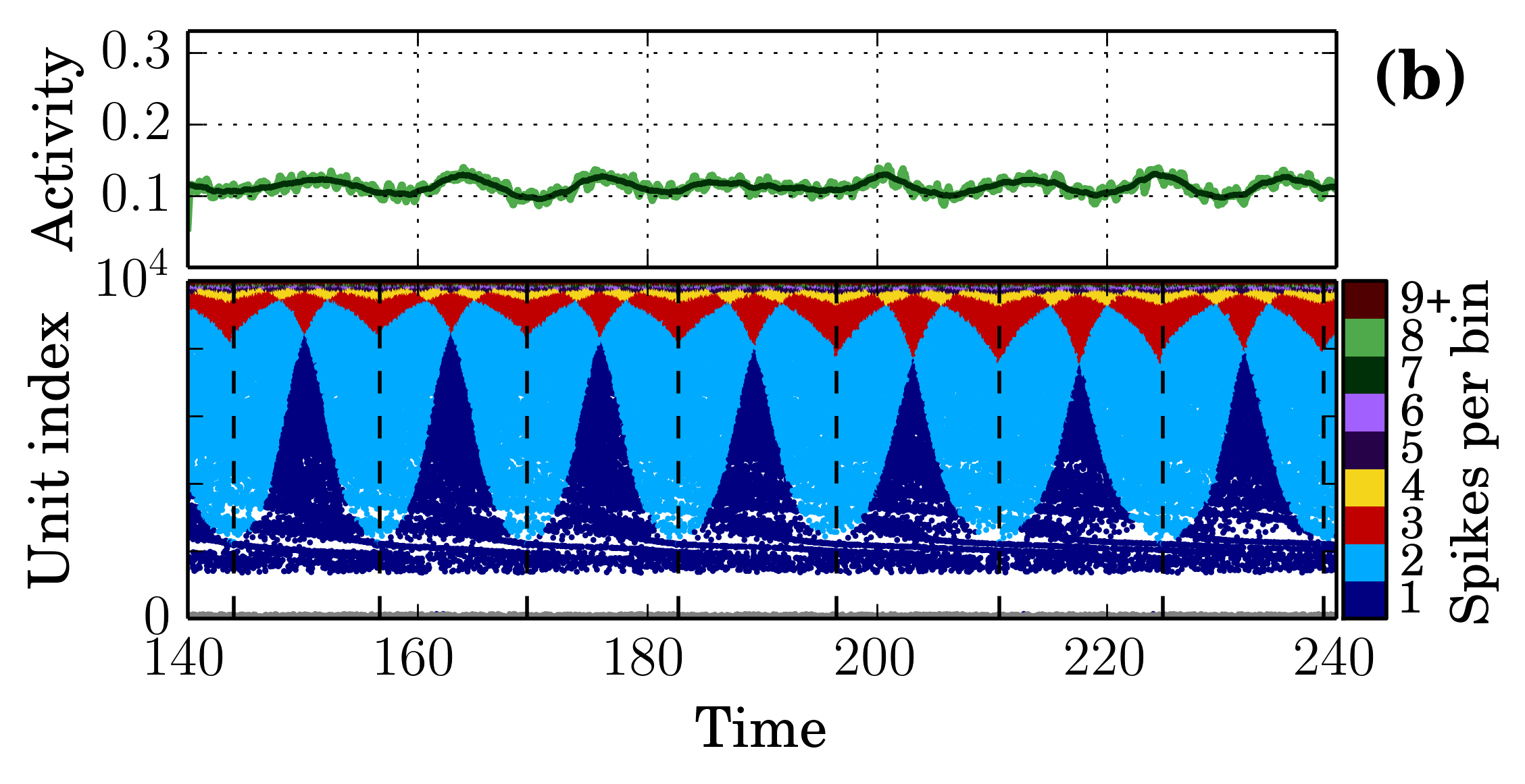}
	
	\includegraphics[width=.65\linewidth]{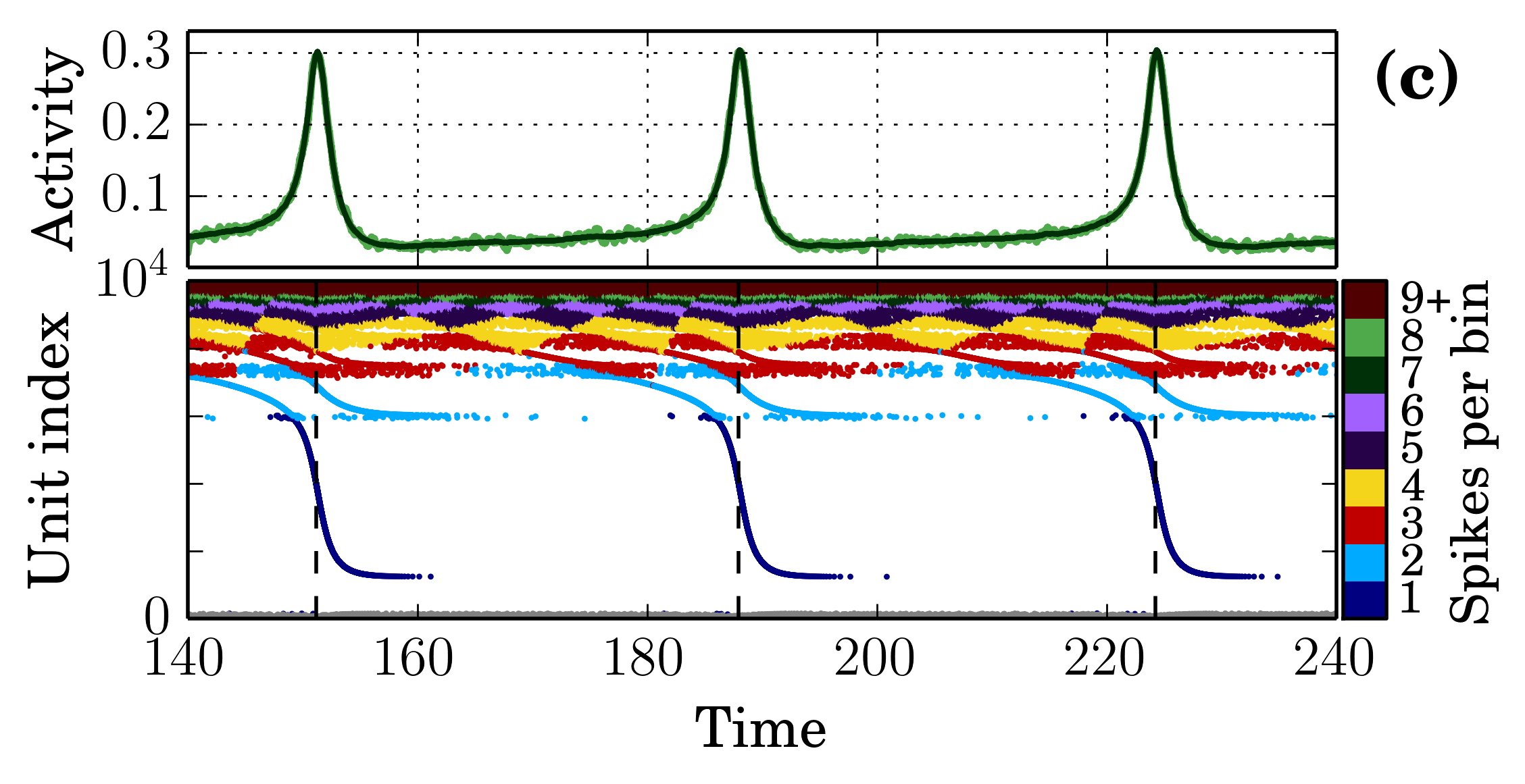}
	\caption{Spiking configuration of the excitatory population at the three qualitatively different activity regimes, with \textbf{(a)} low, \textbf{(b)} high, or \textbf{(c)} oscillating activity, corresponding to the attractors in regions 2, 4 and 3 of Fig.~\ref{fig:bif_Adler} respectively. The upper panel in each subfigure shows the mean activity computed by counting spikes (light green) or by means of Eq.~\eqref{eq:activity} from the order parameters (dark green). The lower panel shows the raster plot of the full population.}
	\label{fig:rasters}
\end{figure}

 Fig.~\ref{fig:sim_attractors} shows the attracting limit sets for the order parameters of simulations of the full network at the representative parameter values chosen in Fig.~\ref{fig:bif_Adler}. These are labeled 1-6, for each of the six regions of the analyzed parameter space. All the simulations are in good agreement with the mean field prediction, although the ``active'' fixed point in regions 4 and 5 presents somewhat large fluctuations with a predominating frequency (see upper panel of Fig.~\ref{fig:rasters}b). Simulations with larger $N$, $\tilde N$ present smaller fluctuations (not shown), suggesting that these are due to finite size effects. Even this behavior can be accounted for by the mean field equation: in these cases, the stable fixed point lies in a slow two dimensional manifold to which all trajectories nearby are rapidly attracted before they spiral into the fixed point, as linearization of the mean field Eq.~\eqref{eq:dzdt} at the fixed point shows. The fixed point has a weak stability along the slow manifold (notice that it is close to losing it at the Hopf bifurcation), and the system is sensitive to fluctuations in those directions. For example, the Jacobian at the fixed point in Region 4 has eigenvalues $\lambda_{1,2}=-0.026\pm0.49i$ associated with the slow manifold, and $\lambda_{3,4}=-0.34\pm0.65i$ associated with the fast one. From the imaginary part of the slow eigenvalues we can correctly predict the period of the fluctuation oscillations, given by $\tau=2\pi/0.49=12.8$. This is in agreement with the one observed in Fig.~\ref{fig:rasters}b. The real parts of the eigenvalues determine the timescale of the transient motions, which are an order of magnitude faster in the fast manifold, supporting the idea of an effective dimensional collapse.

To validate Eq.~\eqref{eq:activity}, we plot in the upper panels of Fig.~\ref{fig:rasters} the mean activity of the excitatory population obtained by counting spikes directly (light green) or by computing it from the order parameters (dark green), as described above. This is done for each of the three qualitatively different regimes found: with low, high, or oscillating activity. The agreement between both methods is impressive. Even in the active fixed point discussed earlier, the activity obtained from the order parameters of the simulation accurately reproduces the fluctuations. These three regimes appear to correspond respectively to the partially synchronous rest state, partially synchronous spiking state and collective periodic wave reported by So et al. in \cite{So2014}.

We can gain further insight on the mechanisms at the level of the individual spikes that generate the distinct macroscopic behaviors, by recording, in the simulation, each unit's spiking times (i.e. the times at which $\theta_i=\pi$). This is somewhat analogous to the experimentalists' raster plots, but for the whole population. Three ``raster plots" are displayed in the lower panels of Fig.~\ref{fig:rasters}, one for each of the regimes studied. In the horizontal axis we represent time, and in the vertical axis we display the unit's index. We ordered the units according to their intrinsic excitability $\omega_i$. The dots represent the individual spikes. Thus, horizontal patterns mean that each unit's behavior depends on its individual excitability, and vertical patterns are associated with synchronization between spikes. To make the synchronization structure clearer, in each case we chose a reference unit that spiked at regular (maximum) time intervals, thereby defining the fundamental frequency of the population. We used the reference unit's spikes, plotted as vertical broken lines, as a natural way to bin time. In this way, each unit in the population fires an integer number of times in each bin, which we use to color-code the spikes. Spikes occurring with $\dot\theta_i<0$ are colored in grey, and we see that only a negligible fraction in the lowest-$\omega$ end of the population does fire backwards in the studied cases. The inhibitory population displays behavior qualitatively similar to the excitatory one in all these cases (not shown).

Fig.~\ref{fig:rasters}a shows that in the fixed point solution, the units either spike non-synchronously, or do not spike at all, depending on whether their intrinsic excitability is above or below a definite threshold. This can be understood by looking at the coupling terms in Eq.~\eqref{eq:model}, which only depend on time through the order parameters. For this reason they present a constant value at the fixed point, say $I_0$. Thus each unit's evolution is ruled by $\dot\theta_i=\omega_i-\cos\theta_i+I_0$, which yields a threshold of $\omega=1+I_0$ below which units do not spike. Above it, they spike with different periods $\tau_i=\int_0^{2\pi} (\omega_i+I_0-\cos\theta)^{-1} {\rm d}\theta$, so that no synchronization can possibly occur at a fixed point for the order parameters. The election of the reference neuron was arbitrary here, since this solution has no natural timescale.

The main differences that Fig.~\ref{fig:rasters}b presents are that a bigger fraction of the population is above threshold, yielding a higher mean activity. Although the majority of the population fires non-synchronously, a number of units tend to synchronize their spikes, causing the mean activity to present fluctuations with regular frequency. Conversely, these fluctuations in the order parameters are needed to make the argument above inapplicable to this case, and to provide a mechanism for synchronization.

In Fig.~\ref{fig:rasters}c, the role that synchronization has in the time dependence of the mean field variables becomes clear: periodic activity peaks are caused by synchronized firing of a large fraction of units within a range of excitabilities. In turn, smaller groups form that synchronize with different spiking ratios 2:1, 3:1, \ldots, 7:1 (notice that each group has a different color). This forms what is called a ``chimera state''. Notably, the mean field variables (the order parameters, and mean activities) do not reflect such a nontrivial pattern of underlying rhythms. Horizontal strips of non-synchronous spikes form at the critical excitabilities at which the firing ratios change, and these units fail to lock to the mean field. The mechanism that synchronizes the different units is the now time-dependent order parameter in the coupling terms of Eq.~\eqref{eq:model}.

We conclude that, in this model, synchronized firing is intimately related to the time dependence of the mean field variables.

\section{Discussion and conclusions} \label{sec:conclusions}

Most mathematicians and physicists who study brain functions use empirical models; simple dynamical systems reflecting one or more important neuro-physiological observations. One celebrated case is the additive Wilson-Cowan empirical model of neural networks. This model is based on the observation that the activity of a neural population increases non-linearly with its input (with the non-linearity reflecting the saturating nature of the response). Remarkably, the simple model that is obtained with coupled excitatory and inhibitory populations is capable of displaying a rich set of dynamical solutions.

Recent advances in the study of coupled oscillators allowed us to obtain equations for variables describing some aspects of the global behavior of the network. In particular, equations were derived for the order parameter of a neural population, describing the degree of synchrony of the solutions. In order to compare a statistical study of a set of coupled oscillators with a phenomenological model as the Wilson-Cowan neural oscillator, it was necessary to go beyond the order parameter and derive the equations for the activity of the network: the average of the actual number of spikes generated over the whole network, at a given time. In this work we performed this calculation, and obtained analytical expressions which could be computed as functions of the order parameters. Two cases were studied in this work: the coupling of Adler units (i.e. elements whose dynamics without coupling were ruled by Adler's equations) and the coupling of theta neurons. In both cases, the couplings were impulsive.

We have found regions of the parameter space where the dynamics of the order parameters derived for the models presented here was equivalent to what is observed in the Wilson-Cowan oscillator. Remarkably, this very simple model is capable of capturing many of the subtle features that a population of coupled units displays after computing its macroscopic behavior from first principles.

\section*{Acknowledgements}

We thank Mat\'ias Leoni, Ana Amador and Gast\'on Giribet for illuminating comments. Funding: This work was supported by CONICET, ANCyT, UBA, and NIH through R01-DC-012859 and R01-DC-006876.

\appendix
\section{On the divergences in the mean activity} \label{app:divergences}

In this appendix we show a method to avoid the divergences in the mean activity (Eq.~\ref{eq:act}) occurring for large values of $\omega$ by slightly changing the distribution function $g(\omega)$. The divergences are due to the slow decay of the Lorentzian function and the term linear in $\omega$ in $v(\omega, \theta, t)$, which means that for large $\omega$, $g\cdot v \sim\omega^{-1}$, whose integral diverges. As it has been said in the main text, a similar behavior but with the opposite sign occurs for large, negative $\omega$, which compensates the divergence giving a finite result.

Thus, the divergence would not occur if $g(\omega)$ were a sharper function. The Ott-Antonsen method contemplates distribution functions having any number of poles off the real axis and being analytical everywhere else. In particular, a slight perturbation of the Lorentzian distribution can be made by taking $g(\omega)$ to be the product of two Lorentzian functions with the same $\omega_0$ but different widths $\Delta$ and $D$, i.e.:

\begin{equation} \label{eq:new_g}
	g(\omega)=\frac{D\Delta(D+\Delta)}{\pi((\omega-\omega_0)^2+\Delta^2)((\omega-\omega_0)^2+D^2)},
\end{equation}

\noindent which has been normalized according to $\int _{-\infty}^\infty g(\omega){\rm d}\omega=1$. The distribution in Eq.~\eqref{eq:new_g} can be made arbitrarily close to the Lorentzian (Eq.~\eqref{eq:lorentzian}) by taking $D$ to be sufficiently large, while at the same time no divergences occur in the integral in Eq.~\eqref{eq:act} for any finite $D$, as now $g\cdot v\sim\omega^{-3}$ for large $\omega$. Although, as we show below, the dynamics of the order parameters now becomes 8-dimensional, it is logical to expect that it will converge to the 4-dimensional one described in Eq.~\eqref{eq:dzdt} for sufficiently large $D$. We will now make this statement quantitative.

With the new distribution function, all steps in the main text remain valid until Eq.~\eqref{eq:z_alpha}, which, introducing Eq.~\eqref{eq:new_g} in Eq.~\eqref{eq:z_int_galpha} and integrating by residues, now becomes:

\begin{equation*} \label{eq:new_z}
	z(t)=(1+\mu)z_1(t)-\mu z_2(t),
\end{equation*}

\noindent where we have defined the two new complex quantities $z_1(t)=\alpha(\omega_0+i\Delta,t)$ and $z_2(t)=\alpha(\omega_0+iD,t)$, and the perturbation parameter $\mu=\Delta/(D-\Delta)$. We also define the analogous quantities $\tilde z_1(t)$, $\tilde z_2(t)$, $\tilde \mu$, etc. for the inhibitory population. In the limit $\mu\to0$ the two distributions Eqs.~\eqref{eq:lorentzian} and \eqref{eq:new_g} become identical, and $z_1(t)\to z(t)$. In the general case, evaluation of Eq.~\eqref{eq:dalphadt} at $\omega=\omega_0+i\Delta$ and $\omega=\omega_0+iD$ for each of the two populations yields the 8-dimensional mean-field dynamics:

\begin{subequations} \label{eq:8d}
	\begin{align}
		 \dot z_1 =& \big[ -\Delta + i \big( \omega_0 + I(z,\tilde z)\big) \big] z_1 - \mfrac{i}{2} \left( 1+z_1^2 \right) \label{eq:8d_z1E} \\
		 \dot{\tilde z}_1 =& \big[ -\tilde\Delta + i \big( \tilde\omega_0 + \tilde I(z,\tilde z) \big) \big] \tilde z_1 - \mfrac{i}{2} \left( 1+\tilde z_1^2 \right) \label{eq:8d_z1I} \\
		 \dot z_2 =& \big[ -D + i \big( \omega_0 + I(z,\tilde z)\big) \big] z_2 - \mfrac{i}{2} \left( 1+z_2^2 \right) \label{eq:8d_z2E} \\
		 \dot{\tilde z}_2 =& \big[ -\tilde D + i \big( \tilde\omega_0 + \tilde I(z,\tilde z) \big) \big] \tilde z_2 - \mfrac{i}{2} \left( 1+\tilde z_2^2 \right), \label{eq:8d_z2I}
	\end{align}
\end{subequations}

\noindent where

\begin{equation*}
	\begin{split}
		I(z,\tilde z) &= k_E(1-(1+\mu)\text{Re\,} z_1 + \mu \text{Re\,} z_2 ) - k_I (1-(1+\tilde\mu)\text{Re\,} \tilde z_1 + \tilde\mu \text{Re\,} \tilde z_2), \\
		\tilde I(z,\tilde z) &= \tilde k_E(1-(1+\mu)\text{Re\,} z_1 + \mu \text{Re\,} z_2) - \tilde k_I (1-(1+\tilde\mu)\text{Re\,} \tilde z_1 + \tilde\mu \text{Re\,} \tilde z_2 ),
	\end{split}
\end{equation*}

\noindent couple all the equations.

For large $D$, $z_1$ and $\tilde z_1$ become the only active degrees of freedom; departures from the 4-dimensional dynamics Eq.~\eqref{eq:dzdt} are represented in coupling terms with $z_2$, $\tilde z_2$, of the form $k\mu\text{Re\,} z_2$. These perturbation terms are small not only because they are weighed by the small parameters $\mu, \tilde\mu$, but also because $z_2$ and $\tilde z_2$ are themselves small, as can be seen by taking the radial component of Eqs.~\eqref{eq:8d_z2E} and \eqref{eq:8d_z2I}. Letting $z_2=\rho_2 e^{i\psi_2}$, then $\dot\rho_2=\text{Re\,}\{\dot z_2 e^{-i\psi_2} \}$ is bounded from above:

\begin{equation*}
	\dot\rho_2 = -D\rho_2-\mfrac{1}{2}(1-\rho_2^2)\sin\psi_2\leqslant -D\rho_2 +\frac{1}{2}.
\end{equation*}

\noindent Thus, in the stationary regime $\rho_2\leqslant(2D)^{-1}$, as $\dot\rho_2<0$ for larger $\rho_2$, and similarly $\tilde\rho_2\leqslant(2\tilde D)^{-1}$. The perturbation terms are then bounded by $|k \mu \text{Re\,} z_2|\leqslant k\Delta/((D-\Delta)2D)\approx k\Delta/(2D^2)$. Taking $D,\tilde D\gg1\gg\Delta, \tilde\Delta$, all the perturbation terms quickly become negligible.

\section{Dimensional collapse for symmetric parameters} \label{app:2D}

In this appendix we show analytically that $z=\tilde z$ is a (two dimensional) invariant, stable manifold in the special case that the system's parameters are chosen symmetrically for both populations. This degenerate election should be considered as a convenient ``attack point'' to Eqs.~\eqref{eq:dzdt}, since the existence of a two dimensional invariant stable manifold should be robust under variations in the parameters up to some extent.

We first rewrite Eqs.~\eqref{eq:dzdt} in terms of the new variables $z_\pm\equiv (z\pm\tilde z)/2$, which account for the average and the difference between the two populations' order parameters, and evolve according to:

\begin{subequations} \label{eq:dzdt_pm}% zp = z+, zm = z-
	\begin{align}[left = \empheqlbrace\,]
		\begin{split}
			\dot z_+ &= \big[-\Delta_+ + i\big(\omega_{0+} + (k_{E+}-k_{I+})(1-\text{Re\,} z_+) - (k_{E+}+k_{I+})\text{Re\,} z_-\big)\big]z_+\\
			&+ \big[-\Delta_- + i\big(\omega_{0-} + (k_{E-}-k_{I-})(1-\text{Re\,} z_+) - (k_{E-}+k_{I-})\text{Re\,} z_-\big)\big]z_-\\ &-\mfrac{i}{2}\big(1+z_+^2+z_-^2\big),
		\end{split} \label{eq:dzpdt}\\
		\begin{split}
			\dot z_- &= \big[-\Delta_- + i\big(\omega_{0-} + (k_{E-}-k_{I-})(1-\text{Re\,} z_+) - (k_{E-}+k_{I-})\text{Re\,} z_-\big)\big]z_+\\
			&+ \big[-\Delta_+ + i\big(\omega_{0+} + (k_{E+}-k_{I+})(1-\text{Re\,} z_+) - (k_{E+}+k_{I+})\text{Re\,} z_-\big)\big]z_-\\
			&- i z_+ z_-.
		\end{split} \label{eq:dzmdt}
	\end{align}
\end{subequations}

\noindent Here, the parameters have been redefined in an analogous way: $\Delta_\pm = (\Delta \pm \tilde\Delta)/2$, and so on. The symmetric-parameter case corresponds to $\Delta_- = \omega_{0-} = k_{E-} = k_{I-} = 0$, in which the radial component of Eq.~\eqref{eq:dzmdt} reduces to:

\begin{equation*}
	\dot\rho_- = (-\Delta_+ + \text{Im\,} z_+)\rho_-,
\end{equation*}

\noindent with $\rho_- = |z_-|$. Now $z_-=0$ is a solution, which defines the invariant two dimensional manifold $z=\tilde z$. Moreover, it will be the only long term solution for $z_-$ unless $\text{Im\,} z_+>\Delta_+$, at least in some part of its evolution. We continue the argument by ruling this possibility out: if we set $z_-=0$, then the radial part of Eq.~\eqref{eq:dzpdt} reads:

\begin{equation*}
	\dot\rho_+ = -\Delta_+\rho_+ - \frac{1-\rho_+^2}{2 \rho_+}\text{Im\,} z_+,
\end{equation*}

\noindent which is always negative for $\text{Im\,} z_+>0$. Therefore, $z_+$ cannot have solutions exclusively in the upper plane (in particular, fixed points). $z_+$ following a bounded 2-dimensional dynamics, the only other possible attractor with $\text{Im\,} z_+>\Delta_+$ somewhere would be a limit cycle that shrunk towards the origin while in the upper plane and expanded away from it in the lower. However, this last possibility can also be severely constrained by noting that, if $z_-=0$, the evolution Eq.~\eqref{eq:dzpdt} becomes the equation of a single, purely excitatory or inhibitory (depending on the sign of $k_{E+}-k_{I+}$) population, which hardly can oscillate in the parameter range explored in this work
\bibnote{Reports of a population of coupled inhibitory theta neurons presenting oscillatory dynamics exist, but for units in a region of the parameter space where the dynamics are qualitatively different from those of neurons. For example, where the phasors' slowing down would occur at $\theta=\pi$ instead of $0$ \cite{So2014}.}.
Therefore, $z_-=0$ is quite generally a stable invariant manifold for any symmetric choice of the parameters, with a number of fixed points in the $z_+$ lower plane. Numerical simulations support this result.

\begin{figure}
	\centering
	\includegraphics[width=.3\linewidth]{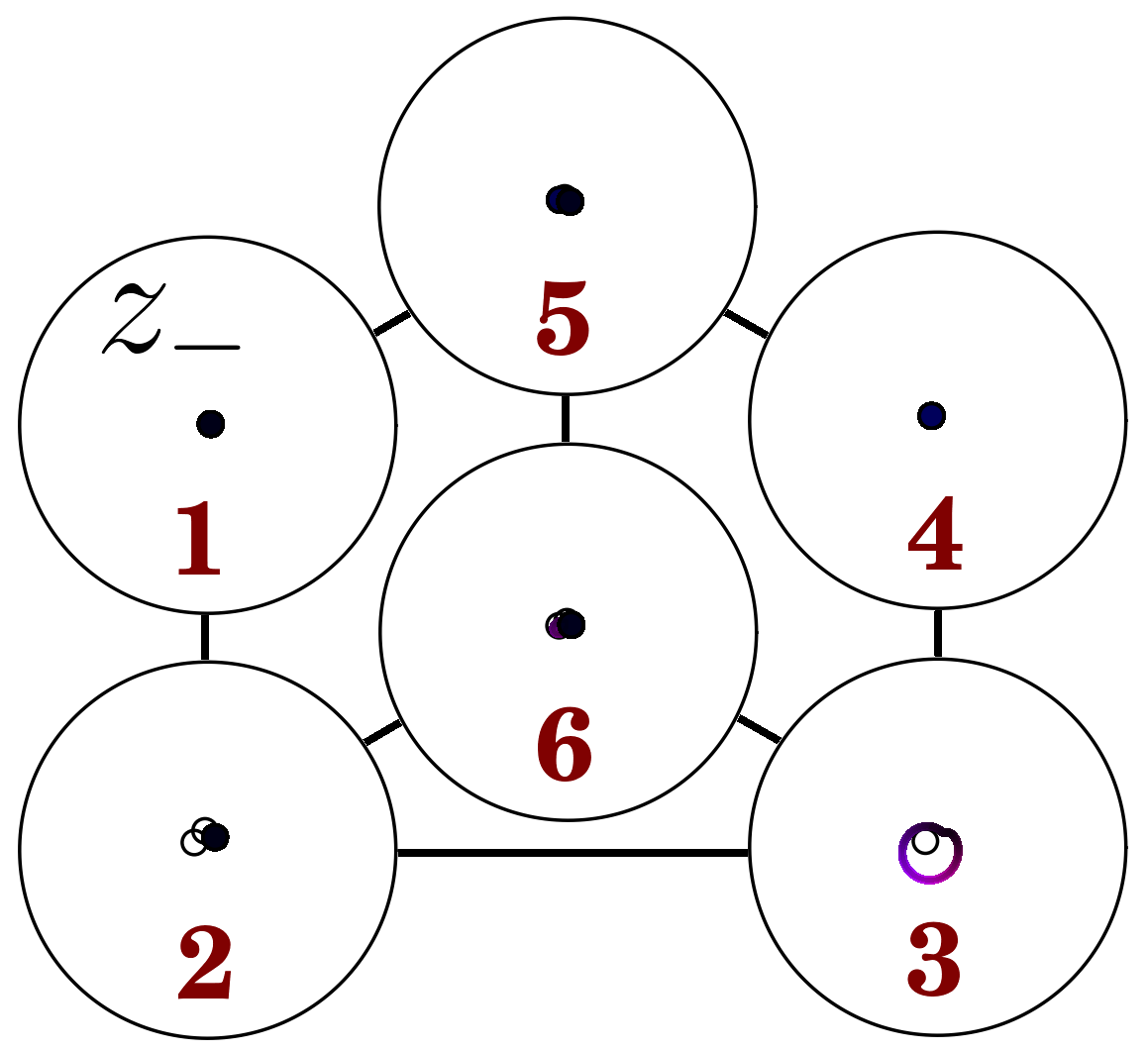}
	\caption{$z_-$ component of the attractor sets for the order parameters of the coupled Adler units model (Eq.~\eqref{eq:dzdt}), at the parameter values labeled 1-6 in Fig.~\ref{fig:bif_Adler}. Filled dots represent stable fixed points, empty dots unstable fixed points and closed curves represent limit cycles.}
	\label{fig:zm}
\end{figure}

Even if no limit cycles exist on the manifold $z=\tilde z$ for the exactly symmetric case, small departures from it in the parameter space can produce rich two-dimensional behavior on the (deforming) stable manifold before a higher dimensionality is explored, as seen in Fig.~\ref{fig:bif_Adler}. In Fig.~\ref{fig:zm} we show the $z_-$ component of the trajectories in the $z,\tilde z$ space (or equivalently, the $z_+, z_-$ space), at the same parameter values 1-6 of the insets of Fig.~\ref{fig:bif_Adler}. Comparing both figures, we see that $z_-$ is much smaller than $z$ (and thus than $\tilde z$ and $z_+$) in the whole region. This supports the idea that the two dimensional manifold to which the dynamics collapse is a deformation of the one defined by $z_-=0$, which would correspond to the exactly symmetric case.

\end{document}